\documentclass{article}[12pt]
\usepackage[cp1251]{inputenc}
\usepackage[T2A]{fontenc}
\usepackage{amsmath,amsfonts,amssymb,amsthm,mathrsfs,bm,color}

\usepackage{graphicx}
\usepackage{cite}

\begin{document}
\vskip 2cm
{\Large 
\centerline{\bf{ ASYMPTOTICS OF A FINITE-ENERGY}}.
 \centerline{\bf{UNIDIRECTIONAL SOLUTION OF THE WAVE EQUATION}}
\centerline{\bf{WITH NON-SPHERICAL-WAVE BEHAVIOR AT INFINITY}}
%\maketitle
\vskip .3cm\vskip .3cm\vskip .3cm
\noindent \centerline{\Large \textbf{Alexandr B. Plachenov}$^1$, and
 \textbf{Aleksei P. Kiselev}$^{2,3}$}}
\vskip .3cm
{\small
\noindent$^1$ \emph{MIREA -- Russian Technological University, Moscow, 119454 Russia}}\\
{\small\noindent$^{2}$ \emph{St. Petersburg Department of V.A. Steklov Mathematical Institute of the Russian Academy of Sciences, St. Petersburg, 191023 Russia;}}\\
{\small\noindent$^{3}$ \emph{Institute for Problems in Mechanical Engineering of the Russian Academy of Sciences,
St. Petersburg, 199178 Russia}}\\
{\small\noindent*e-mail: aleksei.kiselev@gmail.com}

\vskip .5cm\vskip .5cm

\noindent \textbf{Abstract}:
A detailed investigation is presented of a simple unidirectional finite-energy solution of  the 3D wave equation.  Its asymptotics as a spatial point runs to infinity with the wave propagations speed
is a standard spherical wave as  $z<0$, where $z$ is a Cartesian coordinate, and  has an additional factor logarithmic with respect to the distance as $z>0$. Asymptotics for a point running to infinity with an arbitrary constant speed is  discussed.

\vskip .3cm\vskip .3cm

\pagebreak

 \section{ Introduction}This note is devoted, as V. I. Arnold would have contemptuously put it  \cite{Arn}, to a particular property of a particular solution of a particular equation.
  Indeed,   we are concerned with a thorough investigation of the asymptotic behavior at infinity of a simple explicit solution  of the 3D wave equation
\begin{equation}\label{we}
c^2(u_{xx}+u_{yy}+u_{zz})-u_{tt}=0\,,\,\,\, c={\text{ const}} >0,
\end{equation}
in $\mathbb{R}^3\times\mathbb{R}^1$. The study of this solution\footnote{impatient reader may look at the expression \eqref{int=} }
    pushes us to consider novel objects and introduce corresponding novel definitions. We hope that our work will give an impetus to the study of the properties of a wide class of solutions to hyperbolic (and perhaps ultrahyperbolic) equations.

 Energy of the solutions under consideration is finite but its far-field large-time asymptotics
    at $R=\sqrt{x^2+y^2+z^2} \approx ct$
    has  a form of a classical spherical wave only at $z<0$. This property is closely related to the fact that our solution is unidirectional, i.e., all its plane-wave constituents have the
    speeds with non-negative projections on the $z$ axis. Such solutions are currently attracting attention in optics research (see, e.g., \cite{Zam,PA,Lekner18,Biru,BeSa23,SPK24,PSK25}, a fresh review is given in \cite{PlaDD}).
    Additionally, in Section 6 we study the asymptotics of the solution
    at $R \approx \kappa ct$ with $\kappa\ne1$, which we call
 the Demchenko-type asymptotics.
    We also mention modifications of the solution under consideration  that have similar
 non-standard asymptotics within a cone with an arbitrary opening angle $\chi_0$,\, $0<\chi_0<\pi$.

 It should be noted that a function similar to ours was simultaneously touched upon in an interesting work \cite{25}.
  We are commenting on this work in Section \ref{CON}.

\section{Far-field asymptotics at large values of time. Classical and nonclassical running }

 In mathematical physics are often encountered solutions of the equation \eqref{we}
 that behave at infinity as a diverging spherical wave (see, e.g., \cite{Frie,LaPh,Bla,MoPr,Pla} among many others).  This means that with a classical running to infinity  of a point ${\bf R}=(x,y,z)\in\mathbb{R}^3$ in a given direction ${\bf n}$ with the speed $c$, see \eqref{we}, the asymptotics is
 \begin{equation}\label{Sph}
u\approx \frac{F(s, {\bf n})}{R},\, R\to\infty, \, t\to +\infty\,
\end{equation}
 where
\begin{equation}\label{s}
s=R-ct\,.
\end{equation}
Here, $R=|{\bf R}|=\sqrt{x^2+y^2+z^2}$,
    and
${\bf n}={\bf R}/{ R}$ is the unit vector of the direction in which the point $(x,y,z)$ runs to infinity, and $s$ is assumed bounded. In other words, under such a  running in the direction ${\bf n}$, the  solution tends to  the right-hand side of \eqref{Sph}.

The function $F$  known as \emph{pattern} (and also as directivity, or diagram, etc.), is defined   as the limit under the consistent growth of $R$ and $t$,
    i.e., with $R-ct=s$ bounded:
\begin{equation}\label{BMP}
  F(s, {\bf n})=\left. \lim_{t\to +\infty}[ct {\cdot} u({\bf R},t)]\right|_{R=s+ct}.
\end{equation}
It is useful in obtaining
    representations suitable for analysis of solutions  \cite{Frie,Bla,MoPr,Pla,PK2}.
Uniform convergence with respect to  $s\in \mathbb{R}$ is not assumed.
Neither it is assumed with respect to  angular variable ${\bf n}$.
    The assumption that  $R\approx ct$, allows us to write \eqref{BMP} as
    \begin{equation}\label{PBP-R}
    F(s, {\bf n})=\left. \lim_{R\to+\infty}[R {\cdot} u({\bf R},t)]\right|_{t=(R-s)/c}.
    \end{equation}

    If the limit \eqref{BMP} exists for any direction ${\bf n} \in S^2$
    ($S^2$ is the unit sphere), then we call it a \emph{global pattern}.
If the limit \eqref{BMP} exists not for all directions ${\bf n}$,  then we call it a \emph{local pattern}.

A solution of \eqref{we} is conveniently characterized by the initial data
 \begin{equation}
 u({\bf R},t)|_{t=0}=u_0({\bf R})\,,\,\, u_t({\bf R},t)|_{t=0}=u_1({\bf R})\,,
 \end{equation}
where a particular time instant $t=0$ can be replaced by any other one.
Some sufficient conditions for the existence of a global pattern in terms of the decreasing of  the Cauchy data
$u_0({\bf R})$ and $u_1({\bf R})$ as $R\to\infty$  are listed, e.g., in \cite{PKDU}.
These conditions are stronger than the finiteness of energy
\begin{equation}\label{E=}
  \frac{1}{2}\iiint_{\mathbb{R}^3} \left(|\nabla u|^2+\frac{1}{c^2}\left|u_t\right|^2\right)dx\,dy\,dz<\infty.
\end{equation}

 Our recent paper \cite{PKDU} have provided
 a simple example of a function with finite energy, for which
 $$u \left(
 {\bf R},\frac{R-s}{c}
 \right)
 =O\left(\frac{\ln R}{R}\right).$$
 for all ${\bf n}\in S^2$.
 Here, we discuss a finite-energy
  solution of \eqref{we}  having a local pattern for some directions but for others having an additional factor $\ln R$.

   Together with the  above classical
   asymptotics with bounded $s$, see \eqref{s}, we investigate the
    asymptotics as a point runs
    to infinity with a constant speed different from $c$,
\begin{equation}\label{s+}
 R-\kappa ct=\sigma
\end{equation}
with bounded $\sigma\in \mathbb{R}$.
    We call the following nonclassical-running asymptotics
\begin{equation}\label{BMP+}
 F^\kappa(\sigma,{\bf n})=
\left. \lim_{t\to+\infty}[ct {\cdot} u({\bf R},t)]\right|_{R=\sigma+\kappa ct}\,.
\end{equation}
  the  \emph{Demchenko-type
    asymptotics}.\footnote{As far as we know,
    such an asymptotics was first addressed by M. N. Demchenko
    in \cite{Dem} where solutions of the Klein-Gordon-Fock equation were considered. }
In what follows, we confine ourselves to
\begin{equation}\label{k>0}
\kappa >0.
\end{equation}

    If for a given $\kappa$ the limit \eqref{BMP+} exists for any direction ${\bf n} \in S^2$,
  then we call it a \emph{global  $\kappa$-pattern}. If it exists not for all direction, we call it a \emph{local $\kappa$-pattern}.
In Section 6 we establish that for  each $\kappa\ne1$ the function under consideration has nonzero global  $\kappa$-pattern,
    and it
    has the asymptotic form
         \begin{equation}\label{Sphkappa}
u\approx
 \frac{F^\kappa(\sigma, {\bf n})}{R},\, R\to\infty, \, t\to +\infty\,.
\end{equation}
 This pattern appears to be independent on $\sigma$.

 Let us describe  the solution, 
 the study of which is the purpose of this work.
 
\section{The So function}
Consider first the auxiliary function
\begin{equation}\label{So}
      v=\frac{1}{(z_*-\textsf{S})\textsf{S}}\,,
    \end{equation}
where
\begin{equation}\label{SoS}
    \textsf{S}=
 {\textsf{S}(\textbf{R},t)}=   \sqrt{c^2t_*^2-\rho^2}\,,
    \end{equation}
    $$z_*=z+i\zeta\,, \,\,t_*=t+i\tau\,,$$
 $\rho=\sqrt{x^2+y^2}$, $\zeta$ and $\tau>0$ are free real parameters subject to constraint
\begin{equation}\label{ze<cta}
    \zeta<c\tau,
    \end{equation}
    which guarantees the absence of singularities as the root branch is chosen so that $\textsf{S}|_{x=y=0} = ct_*$.

The function \eqref{So}  which we call \emph{the So function} in honor of our co-author Irina  So, was introduced in \cite{SPK20os,SPK20} as a simple model of a ultrashort low-cycle optical pulse. The expression \eqref{So} is a the most simple
of finite-energy unidirectional solutions of the wave equation. For different relationships between free parameters, the So function can model pancake-like, ball-like and needle-like pulses \cite{SPK20,PSK25}.

As shown in \cite{SPK20}, the function  $v=v({\bf R},t)$ satisfies
the wave equation  \eqref{we} in $\mathbb{R}^3\times\mathbb{R}$,  $v\in L_2(\mathbb{R}^3)$. In \cite{BeSa23} it is proved that it is unidirectional, in \cite{PK2} its relations with spherical waves was described, in \cite{SPK24} its modifications were employed for modeling electromagnetic pulses. Modification of \eqref{So}
 whose pattern is localized in arbitrary cone was considered in \cite{PSK25}.

\section{Antiderivative of the So function}

The solution we are starting to explore arose when searching for simple expressions for the components of the Hertz's vector of electromagnetic fields \cite{SPK24}.
Let
 \begin{equation}\label{int}
 u(\textbf{R},t)=c\int_{-\infty}^{t}v(\textbf{R},t')dt'.
    \end{equation}
The result of a bit long though   elementary calculation, is:
 \begin{equation}
 \begin{gathered}\label{int=}
 u(\textbf{R},t) =  \frac{1}{\sqrt{z_*^2+\rho^2}} \ln \frac{ct_*+\textsf{S}-z_*+\sqrt{z_*^2+\rho^2}}{ct_*+\textsf{S}-z_*-\sqrt{z_*^2+\rho^2}}\\
 = \frac{1}{\sqrt{z_*^2+\rho^2}}\ln\frac{P}{Q},
 \end{gathered}
 \end{equation}
with\begin{equation}\label{P}
P=ct_*+\textsf{S}-z_*+\sqrt{z_*^2+\rho^2}\,,\end{equation}
\begin{equation}\label{Q}
Q=ct_*+\textsf{S}-z_*-\sqrt{z_*^2+\rho^2}\,.
\end{equation}
Formula \eqref{int=} can be immediately  verified by differentiation. Indeed,
    \begin{equation}\label{U_t}
      \begin{gathered}
      \frac{U_t}{c}
      = \frac{1}{c\sqrt{z_*^2+\rho^2}}\left\{ \frac{1}{P}-\frac{1}{Q}   \right\}
      \left(c+\textsf{S}_t\right)
      \\=
      \frac{1}{c\sqrt{z_*^2+\rho^2}} \frac{-2\sqrt{z_*^2+\rho^2}}  {(ct_*+\textsf{S}-z_*)^2-z_*^2-\rho^2}  c \left(1+\frac{ct_*}{\textsf{S}} \right)\\
       = \frac{-2 (ct_*+\textsf{S})}{\textsf{S}[(ct_*+\textsf{S})^2-2(ct_*+\textsf{S})z_* -\rho^2]}\\
      = \frac{- (ct_*+\textsf{S})}{\textsf{S}[\textsf{S}(ct_*-\textsf{S})-(ct_*+\textsf{S})z_*]}
      = \frac{1}{(z_*-\textsf{S})\textsf{S}}\,,
      \end{gathered}
    \end{equation}
    which is what was required.

     \textbf{Statement 1}. The function  \eqref{int=} is a solution of the equation \eqref{we}, and it is unidirectional.

     This holds because such is the integrand in \eqref{int}.

     \textbf{Statement 2.} The function \eqref{int=} has finite energy.

    Indeed,  for  $t=0$,
     as is seen from \eqref{So}, $u_t=v=O(R^{-2})$ as $R\to\infty$, whence  $u_t\in L_2(\mathbb{R}^3)$. The first derivatives of \eqref{int=} with respect to spatial variables  admit the estimate $O(R^{-2}\ln R)$ implying $|\nabla u|_{t=0}\in L_2(\mathbb{R}^3)$ which completes the proof.

\section{Asymptotics of  antiderivative of the  So function under the standard point running to infinity}

Consider now the asymptotics of the function \eqref{int=} at $R\to\infty$, $t\to +\infty$
with $s$ in \eqref{BMP}  bounded.

Let $0\leqslant\chi\leqslant\pi$ be the spherical polar angle,
$$\rho=R\sin\chi\,,\,\,\, z=R\cos\chi\,.$$
Obviously,
\begin{equation}
\begin{gathered}
\sqrt{z_*^2+\rho^2}=\sqrt{R^2+2R\cos\chi \cdot i\zeta+O(1)}\\=     R+i\zeta\cos\chi +o(1)\,.
\end{gathered}
\end{equation}
Under the assumption $\cos\chi\neq0$ we have
\begin{equation}
\begin{gathered}
\textsf{S}=\sqrt{R^2\cos^2\chi +2R(ic\tau-s)+O(1)}\\ = R|\cos \chi| + (ic\tau-s)/|\cos \chi|+o(1).
\end{gathered}
\end{equation}

    \subsection{The case $\chi<\pi/2$}
    Consider first the forward half-space where  $\chi<\pi/2$ and thus $|\cos \chi|=\cos \chi$.  Here,
    \begin{equation}\label{aB<}
    \begin{gathered}
     P=     2R+O(1)\,,\\
     Q= \frac{(1+\cos\chi)[i(c\tau-\zeta\cos\chi)-s]}{\cos\chi}+o(1)\,,
     \end{gathered}
     \end{equation}
     and
      \begin{equation} \label{lnR/R}
      u\approx \frac{\ln R}{R}.
      \end{equation}

     \subsection{The case $\chi>\pi/2$}
     For the backward half-space described by $\chi>\pi/2$, $|\cos \chi|=-\cos \chi$. Here,
    \begin{equation}\label{aB>}
      \begin{gathered}
      P= 2R(1-\cos\chi)+O(1)= 2R(1+|\cos\chi|)+O(1)\,,\\
      Q= -2R\cos\chi+O(1)=2R|\cos\chi|+O(1)\,,
      \end{gathered}
    \end{equation}
    whence
    \begin{equation}\label{aB<}
      u\approx \frac{1}{R} \ln \frac{1-\cos\chi}{-\cos\chi} = \frac{1}{R} \ln \frac{1+|\cos\chi|}{|\cos\chi|}.
     \end{equation}
   Expression \eqref{aB<} shows that  $u$ tends to a spherical wave at each value of $\chi$ in the interval $(\frac{\pi}{2},\pi]$, but the tendency is not uniform in $\chi$.

       \subsection{The case $\chi=\pi/2$}
       For the boundary of the forward and backward half-spaces, in the plane $z=0$, $\cos \chi=0$. Here,
    \begin{equation*}
      \begin{gathered}
      \textsf{S}=\sqrt{2R(ic\tau-s)} + o(1)\,,\\
      P= 2R+O(\sqrt{R})\,,\,
      Q= \sqrt{2R(ic\tau-s)}+O(1)\,,
       \end{gathered}
    \end{equation*}
    whence
    \begin{equation} \label{lnR/2R}
      u\approx \frac{\ln R}{2R}.
      \end{equation}

\subsection{Discussion}
The asymptotics \eqref{aB<} is not uniform over the angles,  and the closer the angle is to $\pi/2$, the greater $R$ at which it is valid.

It is worth noting that, unlike the solutions employed earlier for simulation of few-cycle
    optical pulses \cite{Lekner18,Biru,BeSa23,SPK24,SPK20os,SPK20,2D},  for which the  pattern  decreases with the growth of $|s|$, here it is independent of $s$.

 Also, we observe that the leading asymptotic terms for directions of non-standard behavior of the solution, see \eqref{aB<} and \eqref{lnR/2R}, do not depend on $\chi$.

   \section{Demchenko-type nonclassical-running asymptotics of antiderivative of the
   So function}
    Let  the point  ${\bf R}\in \mathbb{R}^3$  run to
    infinity in the direction of vector
      ${\bf n}\in S^2$ with the speed
  $\kappa c$ in such a manner  that $\sigma$ is bounded, see \eqref{s+}.
     We have
      $$R=\sigma+\kappa ct\approx \kappa ct\,,\,\,
      z\approx \kappa ct \cos\chi\,,\,\,$$
      $$\rho \approx \kappa ct  \sin\chi\,,\,\,
      R_*\approx R \approx \kappa ct\,,$$
    and
    \begin{equation}\label{S1}
      \textsf{S} \approx ct \sqrt{1-\kappa^2 \sin^2\chi}.
    \end{equation}
    If the radicand expression in \eqref{S1} is small, the right-hand side here is just
     $\textsf{S}=o(ct)$. If it is negative, the leading term of $\textsf{S}$  is purely imaginary with a positive imaginary part
     (because $\text{Im} \textsf{S}\geqslant c\tau$).

    We address now $P$ and $Q$ carefully monitoring whether they take on small values. It is easy to see that
    \begin{equation}\label{P1}
      P \approx ct(1+\sqrt{1-\kappa^2 \sin^2\chi}-\kappa \cos\chi + \kappa),
    \end{equation}
    where the right-hand side is never small (because $P$
    is either positive, or has a non-vanishing imaginary part).

 Analysis of $Q$ is not so straightforward. At the first glance,
    \begin{equation}\label{Q1}
      Q \approx ct(1+\sqrt{1-\kappa^2 \sin^2\chi}-\kappa \cos\chi - \kappa).
    \end{equation}
    with the error $o(ct)$. However, the right-hand side of \eqref{Q1} vanishes
    when $\kappa=1$ and $\chi\leqslant\pi/2$. This occurs
     on the hemisphere $R=ct$, $z\geqslant 0$, where the
     standard-running
     asymptotics is given by \eqref{lnR/R} or \eqref{lnR/2R}.

     First, consider the solution outside the vicinity of this hemisphere.

     \subsection{The case of non-small $Q$, i.e., $\kappa$ not close to 1, or $\chi>\pi/2$}

   It means that the point is far from the hemisphere $R=ct$, $z\ge0$. There we have
     $$u\approx \frac{1}{R} \ln \frac{1+\sqrt{1-\kappa^2 \sin^2\chi}-\kappa \cos\chi + \kappa}{1+\sqrt{1-\kappa^2 \sin^2\chi}-\kappa \cos\chi - \kappa}\,.$$

     Thus, we showed that the asymptotics of the form \eqref{Sphkappa} holds with
    $$ F^\kappa(\sigma, {\bf n})= \ln \frac{1+\sqrt{1-\kappa^2 \sin^2\chi}-\kappa \cos\chi + \kappa}{1+\sqrt{1-\kappa^2 \sin^2\chi}-\kappa \cos\chi - \kappa}$$
    for $0\leqslant\chi\leqslant\pi$.

    We established that for any $\kappa\ne1$ the function  \eqref{int=} has nonzero global $\kappa$-pattern which is independent of $\sigma$.

    \subsection{The case $\kappa\approx 1$,  $\chi < \pi/2$}

    Consider now $\kappa$ close to $1$, and $\chi < \pi/2$. Obviously, $R\approx ct+ct(\kappa-1)$,
    \begin{equation*}
      P\approx 2ct\approx 2R/\kappa,
      \end{equation*}
      and
    \begin{equation*}
    Q\approx \frac{(1+\cos\chi)[i(c\tau-\zeta\cos\chi) - ct(\kappa-1)]}{\cos\chi},
    \end{equation*}
    Therefore
    \begin{equation}\label{ukappaapprox1}
 u   \approx \frac{1}{R} \ln \frac{2R \cos\chi}{(1+\cos\chi)[i(c\tau-\zeta\cos\chi) - R(\kappa-1)]}.
    \end{equation}
At $\kappa=1$, the leading term of \eqref{ukappaapprox1} coincides with  \eqref{lnR/R}.

\section{Concluding remarks\label{CON}}

This note, together with  \cite{PKDU} and  \cite{25},    
    demonstrates that a solution of the wave equation from an important class of finite-energy functions
    may  have
    an asymptotics different from 
    \eqref{Sph}.
    An interesting discussion of such a behavior of the solution
     is given in \cite{25}. Considering a solution possessing  central symmetry, the authors observe that its non-standard asymptotics  is related to the presence of an incoming spherical wave along with the outgoing one. They suggest that the unusual behavior in other cases has
     a somewhat similar nature. This suggestion seems quite plausible. However,  in the general case (and, in particular, in the case of solution \eqref{int=}) the method of
     splitting
     the field into an incoming and outgoing wave is far from obvious.

    We considered the Demchenko-type asymptotics of the
     antiderivative of the
     So function, and found that the respective nonzero global $\kappa$-pattern
     exists having a singularity only at $\kappa=1$ and $z\geqslant 0$.

As follows from the work \cite{PSK25}, subjecting the function \eqref{int} to  Lorentz transformation with respect to the coordinate $z$, allows   solutions that behave at infinity as $\ln R/R$ inside a cone of a given opening angle
 and have standard asymptotics outside it.
    To be precise, the logarithmic term  arises  in the vicinity of a piece of sphere lying inside the aforementioned cone.
At large $R$ and $t>0$ for which $R=\kappa ct$, $\kappa\ne 1$,  an asymptotics of the form \eqref{Sphkappa} holds both inside and outside the cone.

The above study of a very special solution invites researchers to explore new asymptotic features of finite-energy solutions.

\section*{Acknowledgment}

We are indebted to M. N. Demchenko for a fruitful discussion.

A. K. acknowledges the support of the Ministry of Science and Higher Education of the Russian
Federation through project No. 124040800009-8.

\end{document}